\documentstyle[emulateapj,epsf,apjfonts]{article}

\received{}
\accepted{31 October 2000}

\slugcomment{To appear in The Astronomical Journal, February 2001}

\lefthead{Hrivnak et al.}
\righthead{UV Study of OO Aql}

\begin{document}

\title{An Ultraviolet Study of the Short-Period Binary OO Aquilae}

\author{Bruce J. Hrivnak\altaffilmark{1}}
\affil{Department of Physics and Astronomy, Valparaiso University,
Valparaiso, IN 46383; bruce.hrivnak@valpo.edu}

\author{Edward F. Guinan\altaffilmark{1}, Laurence E. DeWarf, Ignasi Ribas}
\affil{Department of Astronomy and Astrophysics, Villanova University,
Villanova, PA 19085; edward.guinan@villanova.edu,
laurence.dewarf@villanova.edu, iribas@ast.villanova.edu}

\altaffiltext{1}{Guest Observer, {\it International Ultraviolet Explorer}
satellite}

\begin{abstract}

OO Aql is a rare W UMa-type eclipsing binary in which the two solar-type stars
may have only recently evolved into contact. The binary has an unusually high
mass ratio (0.84), and a relatively long orbital period (0$\fd$506) for it
spectral type (mid-G).  Twelve ultraviolet spectra of OO Aql were obtained in
1988 with the {\it IUE} satellite, including a series of consecutive
observations that cover nearly a complete orbital cycle.  Chromospheric
activity is studied by means of the Mg~{\sc ii} h$+$k emission at 2800 {\AA}.
The Mg~{\sc ii} emission is found to vary, even when the emission is normalized
to the adjacent continuum flux.  This variation may be correlated with orbital
phase in the 1988 observations.  It also appears that the normalized Mg~{\sc
ii} emission varies with time, as seen in spectra obtained at two different
epochs in 1988 and when compared with two spectra obtained several years
earlier.  The level of chromospheric activity in OO Aql is less than that of
other W UMa-type binaries of similar colors, but this is attributed to its
early stage of contact binary evolution.  Ultraviolet light curves were
composed from measurements of the ultraviolet continuum in the spectra.  These
were analyzed along with visible light curves of OO Aql to determine the system
parameters.  The large wavelength range in the light curves enabled a
well-constrained fit to a cool spot in the system.
\end{abstract}


\section{Introduction}

The eclipsing binary star system OO Aquilae (HD 187183, $V_{\rm max}=9.2$,
$B-V=+0.76$, mid-G) is an unusual W UMa-type binary because it possesses a mass
ratio near unity (0.84; Hrivnak 1989).  This suggests that the components have
only recently (on astronomical timescales) come into contact, since once in
contact the stars are expected to evolve toward low mass ratio (Webbink 1976; 
Vilhu 1982).
The orbital period of 0$\fd$506 is long compared to other G spectral types
stars with W UMa-type light curves.  This results in an unusually large angular
momentum for a cool W UMa-type system, again suggesting that it has only
recently evolved into contact (Mochnacki 1981).  Thus OO Aql represents a rare,
transient phase in the evolution of contact binary star systems.  On the basis
of his modern radial velocity study using the cross-correlation technique,
Hrivnak (1989) carried out a consistent analysis of both the light and velocity
curves of OO Aql.  He determined the absolute parameters of the two stars with
high precision (M$_1$=1.04$\pm$0.02 M$_{\sun}$, R$_1$=1.39$\pm$0.02 R$_{\sun}$,
M$_2$=0.88$\pm$0.02 M$_{\sun}$, R$_2$=1.29$\pm$0.02 R$_{\sun}$).  By comparing
its properties with stellar models, Hrivnak presented evidence that the stars
have an age of $\sim$8 Gyr.  Thus OO Aql
consists of a pair of mid-G stars in contact, which appear to be somewhat more
evolved than the Sun.  It falls into the subclass of A-type W UMa-type
binaries, in which the more massive component is eclipsed at primary minimum;
this is unusual among the cooler (G and K spectral type) W UMa-type binaries.

W UMa-type binaries of G and K spectral type are known to display a high level
of chromospheric activity, as shown in the study of Rucinski (1985). Most
contact binaries of G$-$K spectral types have periods of 0$\fd$25 to 0$\fd$35
and mass ratios of  0.3 to 0.5.  Thus OO Aql is clearly unusual.  Another
contact binary with a large mass ratio (0.94) is VZ Psc (P=0$\fd$26, K2$-$5).
This system displays a high level of chromospheric activity, with strong and
variable Mg~{\sc ii} h+k and Ca~{\sc ii} H+K emission (Hrivnak, Guinan, \& Lu 1995).  This
suggests that OO Aql might also have variable chromospheric activity.   In OO
Aql, Ca~{\sc ii} emission is not evident in medium-resolution spectra, although there
may be infilling of the broad absorption profiles (Hrivnak 1989).  However,
this apparent absence is most probably the result of a contrast effect, since
OO Aql is hotter and has a much higher continuum in this spectral region
(3900$-$4000 {\AA}) than does VZ Psc.  Since the continuum is lower in the
ultraviolet, the Mg~{\sc ii} h+k emission at 2800 {\AA} provides a better
opportunity to measure the chromospheric activity in OO Aql.  A single
observation of OO Aql made with the low-dispersion LWR camera of the {\it
International Ultraviolet Explorer} ({\it IUE}) satellite by Rucinski (1985) in
1984 indicated an unusually low Mg~{\sc ii} emission level when compared with
other W UMa-type systems.  No x-ray detections, which would measure the coronal
activity, have been reported for OO Aql, not even in the ROSAT Faint Source
Catalogue (Voges et al. 2000).  However, it has been measured as a variable
radio source.  It was one of only a few systems detected in a 3.6-cm radio
continuum survey of W UMa-type binaries, but was not detected at similar
sensitivity when observed two years later (Rucinski 1995).  With this
background of well-determined but unusual properties (compared with other W
UMa-type binary systems), we were motivated to carry out an ultraviolet study
of OO Aql with the {\it IUE} satellite during a complete orbital cycle.  Our
goal was to investigate the level of its Mg~{\sc ii} emission and to search for
phase-related and time-related variations.  As a by-product of these spectra,
we also obtained ultraviolet light curves.  In this paper we present the
results of this study.

\section{{\it IUE} Observations and Reductions}

Observations of OO Aql were made with the {\it IUE} satellite on 1988 August 8
and September 30 (UT).  Nine spectra were obtained on the first date, covering
80\% of the orbit, and three on the second date, which fill in part of the
phase gap and also include a spectrum during primary minimum.  The
long-wavelength primary (LWP) camera, which covers 2000$-$3200 {\AA}, was used
with the large aperture in the low-dispersion mode.  This resulted in a
spectral resolution of $\sim$6 {\AA}.  Exposure times ranged from 18 to 35 min,
and were varied with the phase of the light curve to give approximately
constant signal-to-noise ratios (except for the last observation of the first
date, which was exposed a bit longer).  The observing log is listed in Table 1.
The observing times were converted to heliocentric Julian date (HJD), including
the heliocentric time corrections.  The rising ultraviolet continuum with
numerous absorption blends is clearly seen in these spectra from 2400 to 3000
{\AA}. In each case Mg~{\sc ii} h+k emission is seen rising in the middle of
the broad Mg~{\sc ii} (photospheric) absorption feature.  An example of the LWP
spectrum of OO Aql in this region is shown in Figure 1.

The primary goal of the observations was to obtain good spectra in the region
around the chromospheric Mg~{\sc ii} h+k emission lines at 2800 {\AA}.  At this
low dispersion, the Mg~{\sc ii} doublet (2795.5, 2802.7 {\AA}) cannot be
resolved for these broad-lined stars.  Examples of the Mg~{\sc ii} emission
profiles from spectra taken on the two observing dates are shown in Figure 2.
To measure the strength of the Mg~{\sc ii} emission, one needs to take into
account the underlying Mg~{\sc ii} absorption profile.  One possible approach
to this would be to obtain a similar {\it IUE} spectrum of a mid-G dwarf and
broaden it as necessary to match the spectrum of the binary.  A subtraction of
normalized spectra would then yield a measure of the Mg~{\sc ii} emission.
However, it would be difficult to obtain a spectrum which is an exact match for
OO Aql.  Instead, since the absorption profiles are clearly seen except for the
central emission, one can extrapolate the wings to determine the central
absorption profile in each spectrum with consistency and reasonable accuracy.
This second method was used to determine the underlying absorption profile, and
by subtraction the emission strengths were determined.  Reductions and
measurements were made using the standard NEWSIPS reduction program, but
including the new flux calibration of Massa \& Fitzpatrick (2000).  The
measured values of the emission strengths are listed as f(Mg~{\sc ii}) in Table
1.

Several measurements of the continuum flux density were also made from each LWP
spectrum.  These provide ultraviolet light curves, and also can be used to
investigate variations of the Mg~{\sc ii} emission relative to the ultraviolet
continuum.  To investigate the ultraviolet light curves, the total ultraviolet
flux densities were measured in intervals of 50 {\AA} width, centered at 2575,
2675, and 2975 {\AA}.  These are spectral regions without dominant
absorption features or reseauxs.  Also measured were two regions of 20 {\AA}
width, 2760$-$2780 {\AA} and 2820$-$2840 {\AA}, centered around the Mg~{\sc ii}
feature.  These were later used to determine the Mg~{\sc ii} strength relative to the
adjacent continuum.  All of these measurements are also listed in Table 1.

In addition to obtaining ultraviolet light curves from the flux-calibrated
spectra, one can also use the fine error sensor (FES) on board the satellite to
obtain a visual light curve.  To accomplish this, FES observations of OO Aql
were made immediately before and after most spectra.  The observations each
consisted of two successive measurements with sample times of about 10 sec
each, which were averaged.  To compensate for possible drifts in the
sensitivity of the FES with time, several observations were also made of a
nearby comparison star, HD 186656 ($V=8.70$, $B-V=+0.49$).  Differential FES
magnitudes, transformed approximately to the standard V system using the
calibration of Imhoff \& Wasatonic (1986), are listed in Table 2.  The
estimated uncertainty in the differential V(FES) magnitudes is $\pm$0.015 mag.
For a star of $T_{\rm eff}=6000$ K, the effective wavelength of the broad
bandpass of the FES is 5500 {\AA} (Imhoff 1989).

\section{Analysis of the Ultraviolet and Visible Light Curves}

In Figure 3, the visible (FES) and two ultraviolet light curves of OO Aql are
plotted. The V(FES) light curve shows the typical shape of a W UMa-type binary,
with continuously varying light and curvature in the maxima. 
The orbital phases in Tables 1 and 2 and Figure 3 were calculated using
the results of the period study of this system by Demircan \& Gurol (1996).
Their linear ephemeris was used:
\begin{equation}
\mbox{Min I (HJD)} = 2438239.720 + 0.50678830\;E
\end{equation}
Their sinusoidal ephemeris gave similar results for the times of these 
{\it IUE} observations, producing a phase difference of only $-0\fp002$. 

The ultraviolet light curves are all very similar to one another. The
measurements in each wavelength region were normalized to their maximum 
value and converted to the magnitude scale. As examples, the light
curves at 2675 and 2975~\AA\ are shown in Figure 3. The magnitude scale of the
ultraviolet light curves has been arbitrarily shifted in this figure for 
display purposes.

The shape of a light curve in systems like OO Aql is determined by its orbital
and physical properties, which stay constant over rather long timescales, and
also by transient phenomena, that vary over timescales as short as a few
months, like starspots. As can be seen in Figure 3, the number of measurements 
in both the FES and the ultraviolet light curves is unfortunately too small to 
attempt a complete (orbital, physical, and radiative properties) solution of 
the light curves. Several tests showed that the phase coverage is not dense 
enough (notice the lack of coverage of the minima) for reaching 
convergence. To circumvent the problem, we adopt the orbital and physical
properties of the system from the solution of light curves with better phase
coverage, and then check for consistency with our observations.

Good-quality differential light curves in blue ($B$) and yellow ($V$) passbands
were published by Binnendijk (1968) from 1966 and by Lafta \& Grainger (1985) 
from 1982. Binnendijk's measurements have excellent phase coverage, while Lafta
\& Grainger's observations show a small gap at orbital phases $0.85-0.95$.
They all show the secondary minimum to be nearly as deep as primary, and the
first maximum at phase 0.25 to be somewhat brighter than the second maximum.
We decided to consider both datasets for further analysis because they are both
of good quality.  The analyses for the data sets, however, had to be done
independently, since the observations were made 16 years apart and the effects
of transient phenomena on the light curves, such as starspots, do not allow a
simultaneous and self-consistent analysis.

The light curves were fitted using an improved version of the Wilson-Devinney
program (Wilson \& Devinney 1971; hereafter WD) that includes an atmosphere
model (Kurucz's ATLAS9) routine developed by Milone, Stagg, \& Kurucz (1992)
for the computation of the stellar radiative parameters. A contact
configuration (WD in mode 3) was chosen when running the models. The
bolometric albedo and the gravity brightening coefficients were set to values
of 0.5 and 0.32, respectively, as usually adopted for the convective envelopes
of these late-type stars. The mass ratio ($q\equiv M_2/M_1$) was
fixed to the spectroscopic value of $q=0.843$ and the temperature of the
primary star was set to 5700~K (Hrivnak 1989). The WD program was run in an
iterative mode, in which differential corrections were made to the fitted
parameters until convergence was achieved.  Several sets of starting 
parameters were used in order to explore the full-extent of the parameter 
space and also to make a realistic estimation of the uncertainties.
Simultaneous solutions for the $B$ and $V$ light curves were run for both 
datasets.

The fitted orbital and physical parameters in the light curve analyses were the
orbital inclination ($i$), the temperature of the secondary ($T_2$), the
gravitational potential of the common surface ($\Omega_{1,2}$), the
luminosity of the primary ($L_1$), and a phase offset. In addition, a
dark spot on one of the components was included in the modeling to fit
the asymmetry of the light curves. Both in the case of Binnendijk's and Lafta
\& Grainger's light curves, similarly good fits were obtained with a spot  
about 300~K cooler than the photosphere located on either the primary or
secondary components.  This inability to distinguish between these two
options is not surprising since the difference in wavelength is small between 
the {\it B} and {\it V} light curves and since the spot is visible at 
an orbital phase when both components are in view.
Since the solution to the ultraviolet light curves that we discuss later
clearly indicates that a cool spot is present on the primary component,
we have also assigned the cool spot to the primary component in the
modeling of these visible light curves.
The latitude, the longitude, and the size of the spot were left free in the
differential corrections algorithm. Convergence for the full set of parameters
was reached very quickly for both data sets primarily because of the good phase
coverage. 

The orbital and physical parameters of OO Aql are not expected to change 
between the two observing epochs. Therefore it is pleasantly reassuring to 
find that the best-fitting orbital and physical properties obtained
from the analyses of the two independent datasets are found to agree within
their uncertainties. Only the spot parameters, as one may expect, differ between
the two light curve solutions. In both cases, the latitude\footnote{We are 
listing the latitude rather than the co-latitude, 
which is used internally in the WD program.} of the spot in
the secondary component is found to be around $+40^\circ$, but the spot 
longitudes and sizes are different: a longitude of $60^\circ$ and spot 
radius of $30^\circ$ for Binnendijk's light curves, and a longitude of 
$90^\circ$ and a spot radius of
$65^\circ$ for Lafta \& Grainger's observations. 
In both cases, these result in a second maximum that is fainter than the first.
The fits of the model light curves to the observed normal points are displayed 
in Figure 4.  
The fits are quite good; only slight systematic deviations among the
observations and the synthetic light curves are seen near phase 0.8 in the
light curves of Lafta \& Grainger (both $B$ and $V$). These may be
caused by the simplified spot model (circular cool spot with uniform
temperature) that the WD program uses. The final r.m.s. residuals of the fit 
were about 0.011~mag and 0.012~mag for the light curves of Binnendijk and 
Lafta \& Grainger, respectively, which corroborates their very similar quality. 

We decided to adopt the best-fitting WD solution to Binnendijk's light curves
as our final values, as listed in Table 3. The reason for this is twofold. On
the one hand, no systematic deviations are present in the $O-C$ residuals. On
the other hand, the spot size, and therefore its influence on the light curve,
is smaller in Binnendijk's light curves, which makes the determination of the
intrinsic properties of the stars more reliable. In any case, as we have
already noted, the parameters yielded by the analysis of Lafta \& Grainger's
light curves fall within the uncertainties quoted in Table 3. 

A detailed study of the light curves of Binnendijk has already been carried out
by Hrivnak (1989). We decided, however, to undertake a new analysis chiefly
because of the improved light curve fitting program presently available. The
new modeling has a more sophisticated treatment of the stellar atmospheres with
ATLAS9 Kurucz models (limb darkening, reflection, etc.) and it also includes a
starspot to account for the asymmetry of the light curves. Despite the more
accurate treatment, the values of the best-fitting parameters presented in
Table 3 are in agreement with those already published by Hrivnak (1989),
with the only significant difference being a higher inclination ($90^\circ$) 
in his solution. 
A summary of the physical properties of the two components derived with our 
new solution is presented in Table 4.
These are in agreement with the results of the earlier analysis by 
Hrivnak (1989)\footnote{A 
computational error exists in Hrivnak's (1989) values for 
log~$\it g$ (cgs); they should be 4.18 and 4.17 for the primary and 
secondary components, respectively.}.

Once the orbital and physical properties of the system were determined, the
sparsely-covered FES and ultraviolet light curves could be fit while fixing most
of the parameters. Only the zero-point of the magnitude scale in each of the
light curves and the spot parameters were left free in the differential
corrections algorithm. A total of five light curves (FES and four ultraviolet,
including one at $\overline{\lambda}$ = 2800 \AA\ formed from averaging the
continuum fluxes in the two narrower regions adjacent to the Mg~{\sc ii} feature)
were fitted simultaneously. Interestingly, despite the poor phase-coverage,
convergence of the spot parameters was achieved rapidly and with a high degree
of confidence.  This can be explained in terms of the large wavelength range
(5500 to 2575 \AA) in which the effects of the spot are critical (the contrast
between the spot and the surrounding photosphere increases strongly from the
optical to the ultraviolet). The solution indicates a spot on the primary
component $\sim250$~K cooler than the photosphere located at a latitude of
$+45^\circ$, a longitude of $5^\circ$, and with a radius of $40^\circ$.  Thus
the spot is facing the observer around primary minimum, although it is
partially eclipsed.

Examples of the best-fitting synthetic light curves are shown together with 
the observations in Figure 3. As can be seen, the fits are very good at all
wavelengths and do not show any noticeable systematics, indicating that our
physical model is essentially correct. The r.m.s. residuals are 0.032~mag,
0.041~mag, 0.067~mag, 0.045~mag, and 0.049~mag, for the light curves at
5500~\AA\ (FES), 2975~\AA, 2800~\AA, 2675~\AA, and 2575~\AA, respectively.
The shape of our best-fitting light curves is consistent with the observations
of Demircan, Derman, \& Ekmekci (1991) made around this time (1988--1990), 
i.e. maxima of similar heights and slightly more unequal eclipse depths. 
These contrast with the brighter maximum at phase 0.25 and the slightly more 
similar eclipse depths seen in the light curves of Binnendijk and Lafta \& 
Grainger. The results indicate that the starspot (or major spot group), once
centered around phase 0.75, had moved close to the phase of the primary minimum
when our observations were made.

In Figure 4 a color curve formed from the $\lambda$2675 ultraviolet curve and
the V(FES) light curve is also plotted. The data points represent the 
difference between the observed ultraviolet magnitudes and the magnitudes
interpolated from the best-fitting synthetic V(FES) light curve at the proper
phases. Note the color dependence in the phases around the primary eclipse
despite the similar effective temperature of both components. This color
dependence can be attributed to the increased effect of limb darkening when 
the smaller star transits the larger one (limbs of both stars visible) and to 
the presence of the dark spot partially in view near the phase of the primary 
minimum, in addition to the gravity darkening which reddens both minima.

\section{Analysis of the Mg \sc{ii} Emission}

The values of the Mg~{\sc ii} flux density, as determined by measurements of
the emission above the underlying absorption profiles and listed in Table 1,
indicate that the flux varies.  These are plotted in the top portion of Figure
5 as a function of orbital phase.  It can be seen that there is a clear phase
dependence in the Mg~{\sc ii} strength, with lower values corresponding to
phases of minimum light (eclipses) in the light curve.  Another way to
investigate the presence of a variation in the emission is to simply measure
the total flux within the Mg~{\sc ii} spectral region, relative to the adjacent
continuum.  This was also carried out, and shows a similar phase-dependent
variation in Mg~{\sc ii} emission.

If the Mg~{\sc ii} h+k emission originates from many active regions distributed
around the atmosphere of the two components, then one would expect the net
emission to vary with the projected cross-sectional area of the two stars,
which varies with the orbital phase.   A first-order correction for this effect
can be made by normalizing the Mg~{\sc ii} emission relative to the light
curve.  The two adjacent continuum bands measured from 2760$-$2780 {\AA} and
2820$-$2840 {\AA} were used for this purpose, and the Mg~{\sc ii} emission flux
relative to the average of these was calculated.  This is listed in Table 1 as
R$_{uv}$, where R$_{uv}$ = f(Mg~{\sc ii})/f(uv cont).  Rucinski (1985), in his
study of the Mg~{\sc ii} emission in W UMa-type binaries, normalized the Mg
emission relative to the bolometric flux, R$_{bol}$ = f(Mg~{\sc ii})/f(bol),
where 
\begin{equation}
{\rm f(bol)} = 2.7\times10^{-5} \: 10^{-0.4\;{\rm m(bol)}} \;
\mbox{erg cm$^{-2}$ s$^{-1}$},
\end{equation}
and m(bol) = V + BC.  We also calculated this ratio.  The value of V is
interpolated from the synthetic V(FES) light curve, 
and BC was estimated from the colors
of the system, $(B-V)=+0.76$ (Eggen 1967), and reddening, $E(B-V)=+0.03$ 
(Hrivnak 1989), to be $-$0.15 (Flower 1996).  
Both of these normalized measures of the Mg~{\sc ii} emission show similar effects, 
and in Figure 5 is plotted R$_{uv}$ versus orbital phase.

A variation is seen in the normalized Mg~{\sc ii} emission R$_{uv}$.  Initial
inspection of the bottom portion of Figure 5 suggested a phase dependence, 
since the values based on
the five observations in the phase interval 0.80$-$0.05 are larger than those
of the seven observations in the phase interval 0.05$-$0.70 by 36$\%$ on
average. However, these may instead represent a variation with time, with the 
average R$_{uv}$ value of the three observations from September 1988 larger 
by 35$\%$ than the average R$_{uv}$ value of the nine observations from
August 1988.  The highest Mg~{\sc ii} flux measurement, at phase$=$0.81, may be the
result of chromospheric flaring, which gradually decreased or rotated partially
out of view when the following two observations were made.  Also included in
Table 1 and Figure 4 are the results from single observations of OO Aql made 
four and six  years earlier.  These spectra were obtained from the {\it IUE} 
archives and were reduced and measured in the same way as the others.  
The one from 1982 has a normalized Mg~{\sc ii} flux similar to those of the 1988 
observations, while the one from 1984 (Rucinski) is lower.  
We conclude that the normalized Mg~{\sc ii} flux varies with time, and perhaps it
also varied with phase during our 1988 observations. 
If the normalized Mg~{\sc ii} variation is indeed related to phase in the 1988 
observations, one can investigate its relationship with the visibility of 
the spot.  If we assume that the highest emission value at phase$=0.81$ 
represents a flare and simply look at the other four higher emission
measurements in the bottom of Figure 5, the results seem to be ambiguous.  
The normalized Mg~{\sc ii} emission is high as the spot comes into view 
(phase $=0.85-0.95$), but the spot is partially eclipsed during the high Mg~{\sc ii}
observation at phase$=0.02$, and the observed emission is not high in the 
phase interval $0.05-0.15$.

\section{Summary and Conclusions}

Ultraviolet observations were obtained for the short-period contact binary
OO Aql on two dates in 1988, including nine consecutive observations that
covered most of one orbital cycle.  Mg~{\sc ii} h+k emission is seen in each of
the spectra.  These were measured for emission strength, and also
measurements were made of the continuum at several wavelength intervals 
and used to compose ultraviolet light curves for the binary.  
An analysis of the Mg~{\sc ii} emission indicates that real variations in the
strength of the feature exist for OO Aql, even when normalized
to the surrounding ultraviolet continuum.  These may be correlated with the 
orbital phase, or they may instead indicate overall changes in the 
chromospheric activity level with time.  

Previous studies of phase-related activity in W UMa-type binaries have yielded
similar results.  Eaton (1986) found only a slight phase dependence in the
normalize Mg~{\sc ii} emission for SW Lac (P=0$\fd$32, K0), another contact binary
with a relatively large mass ratio (0.73, Hrivnak 1992).   The study of VZ Psc
(Hrivnak et al. 1995) revealed variations in the normalized Mg~{\sc ii} emission
that were not phase dependent, but that seemed to vary in time.
VW Cep (P=0$\fd$27, G5V+G8V) shows primarily
time-dependent variations in chromospheric Mg~{\sc ii} emission and also in the
transition region emission flux (Guinan \& Gim\'enez 1993).  Thus variations in
the normalized Mg~{\sc ii} emission are common in W UMa-type binaries, but these are
typically not phase dependent, and instead are probably due to variations in
chromospheric activity arising from possible flaring.

General trends have been found for W UMa-type and other short-period binaries
in studies of chromospheric activity versus color or orbital period.  A summary
for W UMa-type binaries has been published by Rucinski (1985), which included
his observation of OO Aql.  While these more recent observations show
OO Aql to have a somewhat higher level of chromospheric activity than at the
time of his 1984 observation, the basic interpretation is the same; OO Aql has
a much lower level of relative activity (by a factor of 2 relative to f(bol))
than do other contact binaries with similar colors.  However, this is
presumably due to the recent evolution of its components into contact. 
While in contact, OO Aql is expected to transfer mass from the less-massive
secondary to the more-massive primary component (Webbink 1976), reducing the
mass ratio and causing the system to evolve to evolve to an F spectral type. 
The level of chromospheric activity in OO Aql is in fact comparable to that 
in W UMa-type systems with the colors of F stars.

Ultraviolet light curves are presented here for the first time for OO Aql.
They show a similar shape and amplitude as the visible light curves.  Since the
two stars have similar temperatures, there is expected 
to be little change in the light curves with wavelength.  
Synthetic light curve solutions were carried out for $B$ and $V$ light
curves for two different epochs, and similar orbital and physical parameters
were obtained.  These were used to model the new ultraviolet light curves,
and good fits were obtained.
The combination of ultraviolet and visible light curves led to a
well-determined fit for a cool spot on the primary component.  It appears 
that the light curves at the different epochs can each be fit by the 
same orbital and physical parameters, with the presence of one major cool 
spot on the primary component that varies in longitude and size with time.
Jeong et al. (1994) carried out a somewhat similar ultraviolet and
visible light curve analysis for SW Lac, also including a single cool spot. 

The high mass ratio and relatively long orbital period lead to a relatively
large orbital angular momentum for OO Aql, which is  larger than that of most
contact binaries but less than that of the short-period detached binaries with
G-type components, such as ER Vul and UV Leo.  It is likely that the detached
systems such as ER Vul (P=0$\fd$70, M$_1$=0.96 M$_{\sun}$, M$_2$=0.89
M$_{\sun}$, Budding \& Zeilik 1987) and UV Leo (P=0$\fd$60, M$_1$=0.99
M$_{\sun}$, M$_2$=0.92 M$_{\sun}$, Popper 1980) will evolve into contact
systems like OO Aql (P=0$\fd$51, M$_1$=1.05 M$_{\sun}$, M$_2$=0.88 M$_{\sun}$) 
through angular momentum loss by magnetic breaking (Mochnacki 1981; Vilhu
1982).  However, ER Vul has a much higher level of variable starspot and 
chromospheric activity than that seen in OO Aql, even though its rotational 
period is longer and its spectral type earlier than those of OO Aql.  
If OO Aql is typical of a system of solar mass stars that have recently 
evolved into contact, then the previous high level of activity must be 
significantly reduced in the contact phase.  Subsequent evolution while in 
contact is expected to decrease the mass ratio, and if this is so, then a 
binary like OO Aql will gradually evolve from G to F spectral type as the
mass is transferred from the secondary to the primary component.  As shown by
Rucinski (1985), the F spectral type systems show a lower activity, consistent
with what is now seen in OO Aql.  

Thus this study of OO Aql has presented an opportunity to learn more about the
properties of a close binary in a rather short-lived stage of its evolution,
shortly after the two stars have evolved into contact.  If OO Aql is typical 
of this stage, then it appears that the level of chromospheric activity must 
be reduced rather quickly upon entering the contact configuration, with the
contact binary continuing to maintain this lower level with some smaller 
variations over time.

\acknowledgments

The assistance of the staff of the {\it IUE} in making these observations and
of the RDAF staff at the Goddard Space Flight Center in carrying out the
preliminary reductions is much appreciated.  We thank John Pritchard for
providing the model atmosphere files for the ultraviolet passbands.  This
research was supported in part by grants from NASA to B.J.H. (NAG 5-964, NAG
5-2645) and to E.F.G. (NAG 5-2160); these are gratefully acknowledged.  I.R.
acknowledges the Catalan Regional Government (CIRIT) for financial support
through a postdoctoral Fulbright fellowship.

\clearpage
\begin{deluxetable}{rrrcrrrrrrrr}
\tablewidth{0pt}
\tiny
\tablecaption{{\it IUE} Observing Log for OO Aquilae
\label{iue_log}}
\tablehead{\colhead{Image No.}
&\colhead{HJD}
&\colhead{Phase}
&\colhead{Exposure} 
&\colhead{f(Mg~{\sc ii})\tablenotemark{a}}
&\colhead{f(2575)\tablenotemark{b}}
&\colhead{f(2675)\tablenotemark{b}}
&\colhead{f(2975)\tablenotemark{b}}
&\colhead{f(2770)\tablenotemark{b}}
&\colhead{f(2830)\tablenotemark{b}}
&\colhead{R$_{uv}$}
&\colhead{R$_{bol}$\tablenotemark{c}}\\
 \colhead{} 
&\colhead{(2,440,000+)}
&\colhead{}
&\colhead{Time(min)} 
&\colhead{}
&\colhead{(50 {\AA})\tablenotemark{d}}
&\colhead{(50 {\AA})\tablenotemark{d}}
&\colhead{(50 {\AA})\tablenotemark{d}}
&\colhead{(20 {\AA})\tablenotemark{d}}
&\colhead{(20 {\AA})\tablenotemark{d}}
&\colhead{}
&\colhead{}}
\startdata
LWP 13801& 7381.601& 0.856& 25&	5.41& 1.60& 2.99& 6.32& 1.28& 1.58& 0.38&10.16\nl
LWP 13802& 7381.647& 0.946& 25&	3.53& 1.14& 1.79& 3.95& 0.72& 0.95& 0.42& 9.43\nl
LWP 13803& 7381.711& 0.073& 28&	3.64& 1.27& 2.32& 5.23& 0.99& 1.29& 0.32& 8.56\nl
LWP 13804& 7381.759& 0.168& 20&	5.01& 1.82& 3.30& 7.19& 1.33& 1.79& 0.32& 8.88\nl
LWP 13805& 7381.817& 0.283& 18&	5.82& 2.07& 3.65& 7.52& 1.50& 1.98& 0.33& 9.58\nl
LWP 13806& 7381.858& 0.362& 20&	4.17& 1.74& 3.33& 6.57& 1.29& 1.65& 0.28& 7.65\nl
LWP 13807& 7381.909& 0.463& 25&	2.82& 1.11& 1.91& 4.07& 0.79& 1.01& 0.31& 8.18\nl
LWP 13808& 7381.958& 0.561& 30&	3.98& 1.35& 2.52& 5.09& 1.09& 1.34& 0.33& 9.65\nl
LWP 13809& 7382.011& 0.664& 35&	5.45& 1.81& 3.54& 7.45& 1.50& 1.92& 0.32& 9.55\nl
LWP 14137& 7434.788& 0.805& 20&	9.05& 1.87& 3.53& 7.16& 1.49& 2.00& 0.52&15.50\nl
LWP 14138& 7434.851& 0.929& 20&	4.34& 1.20& 2.18& 4.88& 0.95& 1.30& 0.39&10.46\nl
LWP 14139& 7434.896& 0.018& 28& 2.94& 0.75& 1.54& 3.19& 0.58& 0.75& 0.44&10.24\nl
\tableline                                                               
LWR 13609& 5155.647& 0.580& 36&	4.09& 1.23& 2.30& 5.22& 0.96& 1.28& 0.37& 8.93\nl
LWP 03362& 5834.767& 0.626& 45&	3.91& 1.63& 3.34& 6.59& 1.34& 1.77& 0.25& 7.34\nl
\enddata
\tablenotetext{a}{Units of 10$^{-13}$ erg cm$^{-2}$ s$^{-1}$.}
\tablenotetext{b}{Units of 10$^{-12}$ erg cm$^{-2}$ s$^{-1}$.}
\tablenotetext{c}{Units of 10$^{-5}$.}
\tablenotetext{d}{Width of integrated flux density interval.}
\end{deluxetable}

\begin{deluxetable}{rrr}
\tablewidth{0pt}
\tablecaption{{\it IUE} FES Light Curve of OO Aquilae (1988)
\label{fes}}
\tablehead{\colhead{HJD} & \colhead{Phase} & \colhead{$\Delta$V(FES)}\\
\colhead{(2,440,000+)}&\colhead{}&\colhead{}}
\startdata
7381.5882& 0.830& +0.67\nl
7381.6125& 0.878& +0.80\nl
7381.6327& 0.918& +0.96\nl
7381.6584& 0.969& +1.36\nl
7381.6959& 0.043& +1.17\nl
7381.7230& 0.096& +0.82\nl
7381.7264& 0.103& +0.79\nl
7381.7462& 0.142& +0.69\nl
7381.7688& 0.187& +0.62\nl
7381.7889& 0.226& +0.57\nl
7381.8264& 0.300& +0.59\nl
7381.8473& 0.342& +0.64\nl
7381.8716& 0.390& +0.74\nl
7381.8945& 0.435& +0.96\nl
7381.9209& 0.487& +1.32\nl
7381.9431& 0.531& +1.15\nl
7381.9931& 0.629& +0.68\nl
7382.0257& 0.694& +0.61\nl
7434.8194& 0.867& +0.70\nl
7434.8409& 0.909& +0.85\nl
7434.8610& 0.949& +1.11\nl
7434.8826& 0.991& +1.51\nl
\enddata
\end{deluxetable}

\begin{deluxetable}{lrr}
\tablewidth{0pt}
\tablecaption{Light Curve Solution for OO~Aquilae \tablenotemark{a}
\label{lc_model}}
\tablehead{\colhead{Parameter}&\colhead{Value}}
\startdata
Inclination (deg)                        & $87.7\pm0.5$    \nl
$\Delta$T$\equiv T_1 - T_2$ (K)          & $+20\pm25$      \nl
$q\equiv{M_2}$/${M_1}$                  & $0.843$\rlap{\tablenotemark{b}}\nl
$({L_2}$/${L_1}$)$_{B}$ 		 & $0.839\pm0.021$ \nl
$({L_2}$/${L_1}$)$_{V}$    		 & $0.844\pm0.018$ \nl
$\Omega_{1,2}$                           & $3.388\pm0.024$ \nl
$f$ (fill-out factor)               & $0.22\pm0.06$\rlap{\tablenotemark{c}}\nl
$r_1$ (volume)                           & $0.415\pm0.004$ \nl
$r_2$ (volume)                           & $0.385\pm0.004$ \nl
\enddata
\tablecomments{The uncertainties listed are standard deviations
derived from a sampling of the $\chi^2$ minimum.}
\tablenotetext{a}{Based on the solution obtained from the analysis of the
light curves of Binnendijk (1968).}
\tablenotetext{b}{Fixed at the spectroscopic value of Hrivnak (1989).}
\tablenotetext{c}{$f\equiv(\Omega_{1,2}-\Omega_{\rm inner})/
(\Omega_{\rm outer}-\Omega_{\rm inner})$.}
\end{deluxetable}

\begin{deluxetable}{lrr}
\tablewidth{0pt}
\tablecaption{Physical Properties of the Components of OO~Aquilae
\label{star_properties}}
\tablehead{\colhead{}&\colhead{Primary}&\colhead{Secondary}}
\startdata
Mass (M$_{\odot}$)                       & $1.05\pm0.02$ & $0.88\pm0.02$ \nl
Radius (R$_{\odot}$)                     & $1.38\pm0.02$ & $1.28\pm0.02$ \nl
$\log g$ (cgs)                           & $4.18\pm0.01$ & $4.17\pm0.01$ \nl
$T_{\rm eff}$ (K)                        & $5700\pm300$  & $5680\pm300$  \nl
$\log (L/L_{\odot})$                     & $0.26\pm0.09$ & $0.19\pm0.09$ \nl
$M_{\rm bol}$ (mag)                      & $4.11\pm0.22$ & $4.29\pm0.22$ \nl
$M_{\rm v}$ (mag)                        & $4.21\pm0.22$ & $4.39\pm0.22$ \nl
\enddata
\tablecomments{Adopted the radial velocity curve solution of Hrivnak (1989);
BC from Flower (1996).}
\end{deluxetable}

\clearpage

\begin{figure*}
\plotone{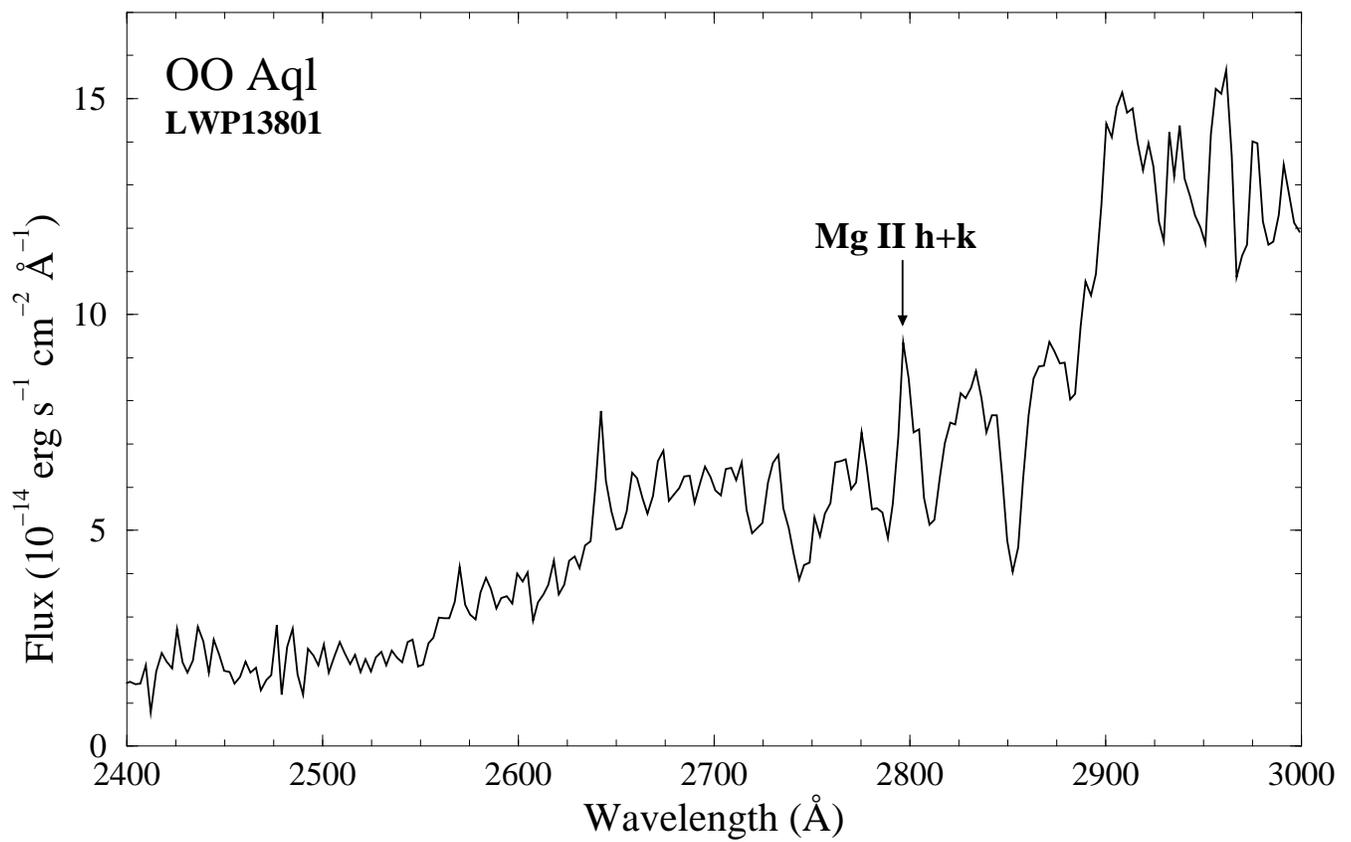}
\figcaption[]
{A sample {\it IUE} spectrum of OO Aql, showing the continuum from 
2400 to 3000 {\AA} with the Mg~{\sc ii} emission marked.
\label{f1.eps}}
\end{figure*}

\begin{figure*}
\plotone{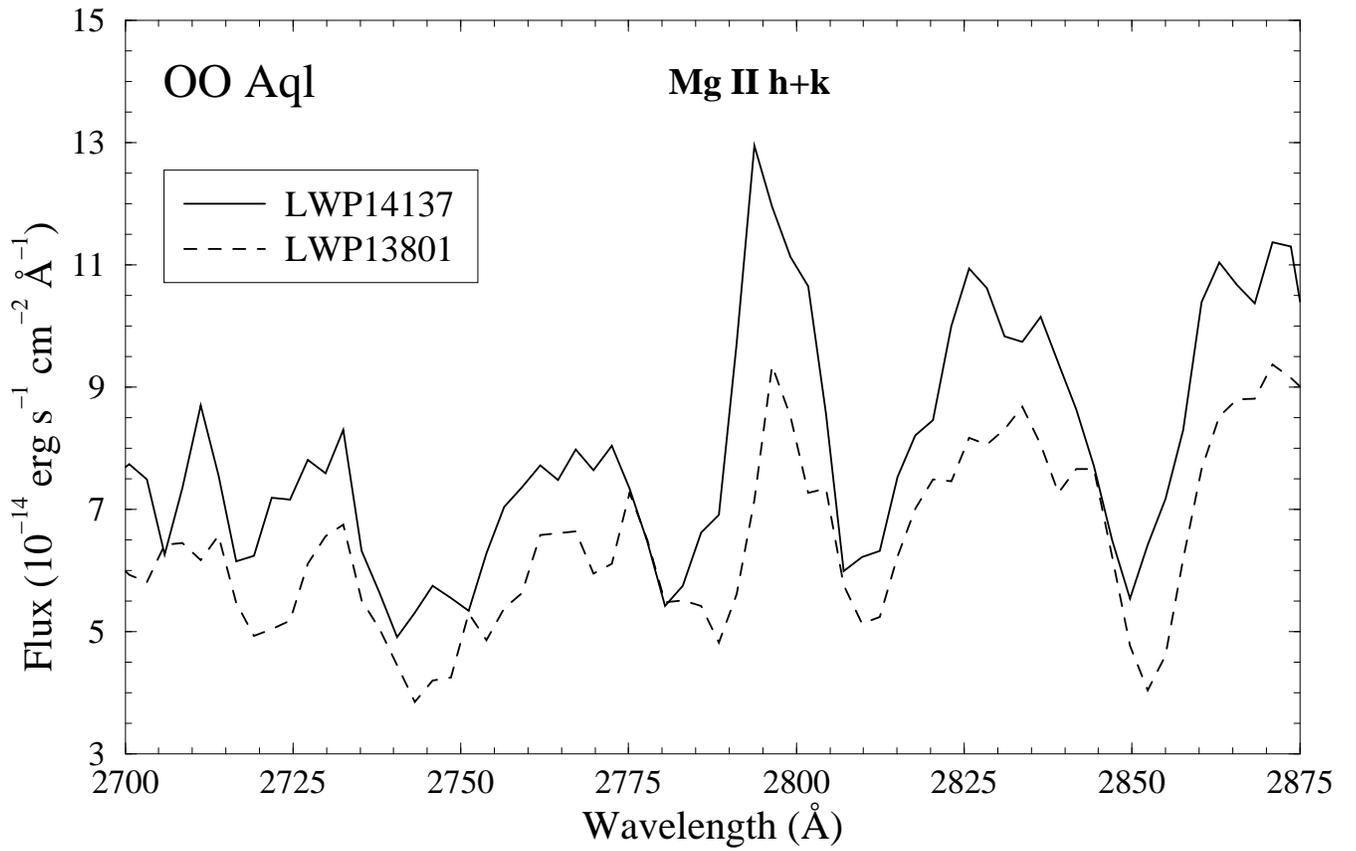}
\figcaption[]
{Spectra of OO Aql in the region of the Mg~{\sc ii} feature, obtained around 
the same phase but on different dates 
(LWP 13801, $0\fp85$; LWP 14137, $0\fp81$).   Note the difference in the 
emission strength between the two spectra. 
\label{f2.eps}}
\end{figure*}

\begin{figure*}
\plotone{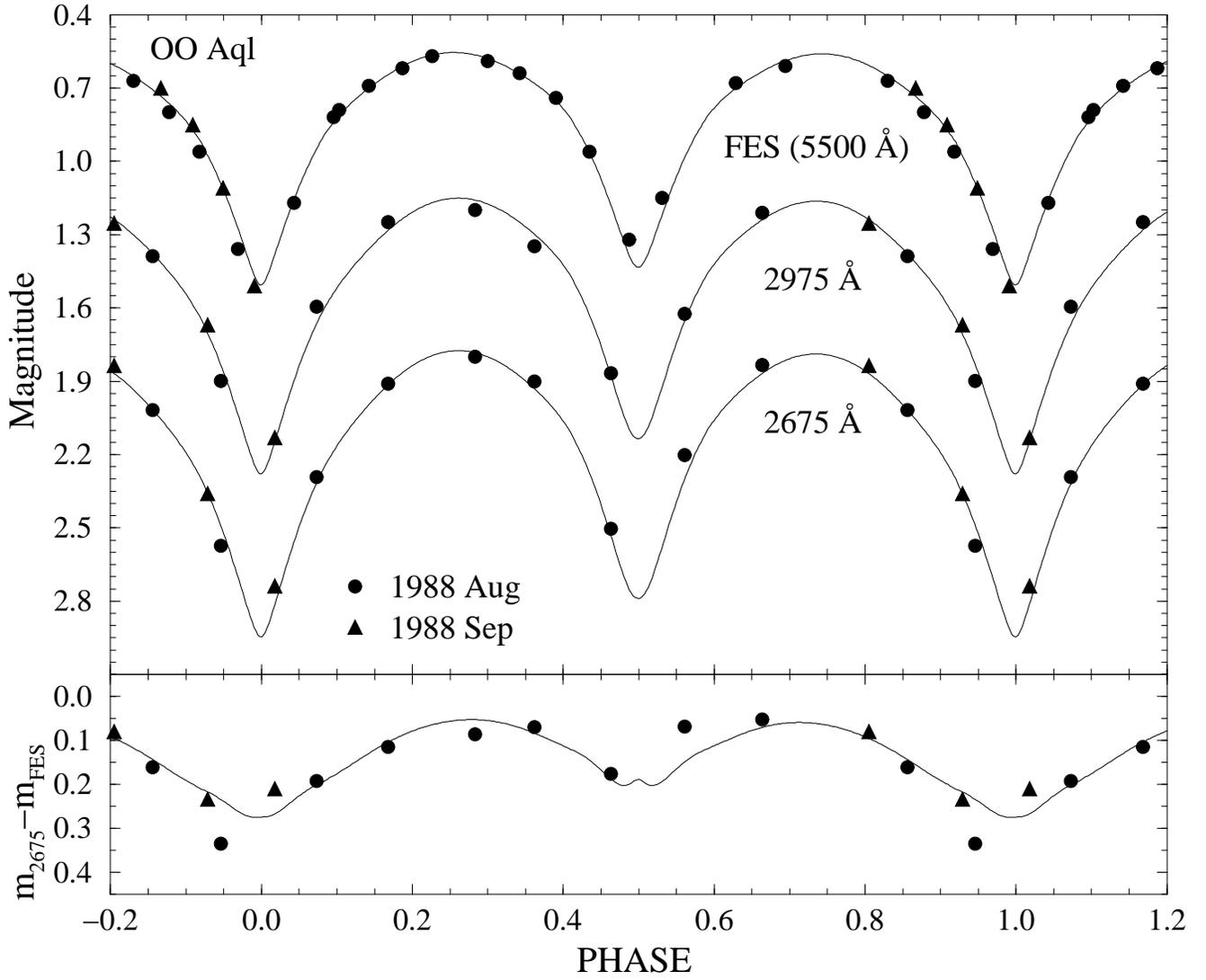}
\figcaption[]
{(upper panel) Visible (FES) and representative ultraviolet light curves of 
OO Aql, and 
(lower panel) an ultraviolet minus visible color curve formed from the 
lower and upper light curves of the upper panel. 
Different symbols are used to distinguish the different dates: 1988
Aug (filled circles), 1988 Sep (filled triangles). The magnitude scale of the
ultraviolet light curves has been arbitrarily shifted for display purposes.
\label{f3.eps}}
\end{figure*}

\begin{figure*}
\plotone{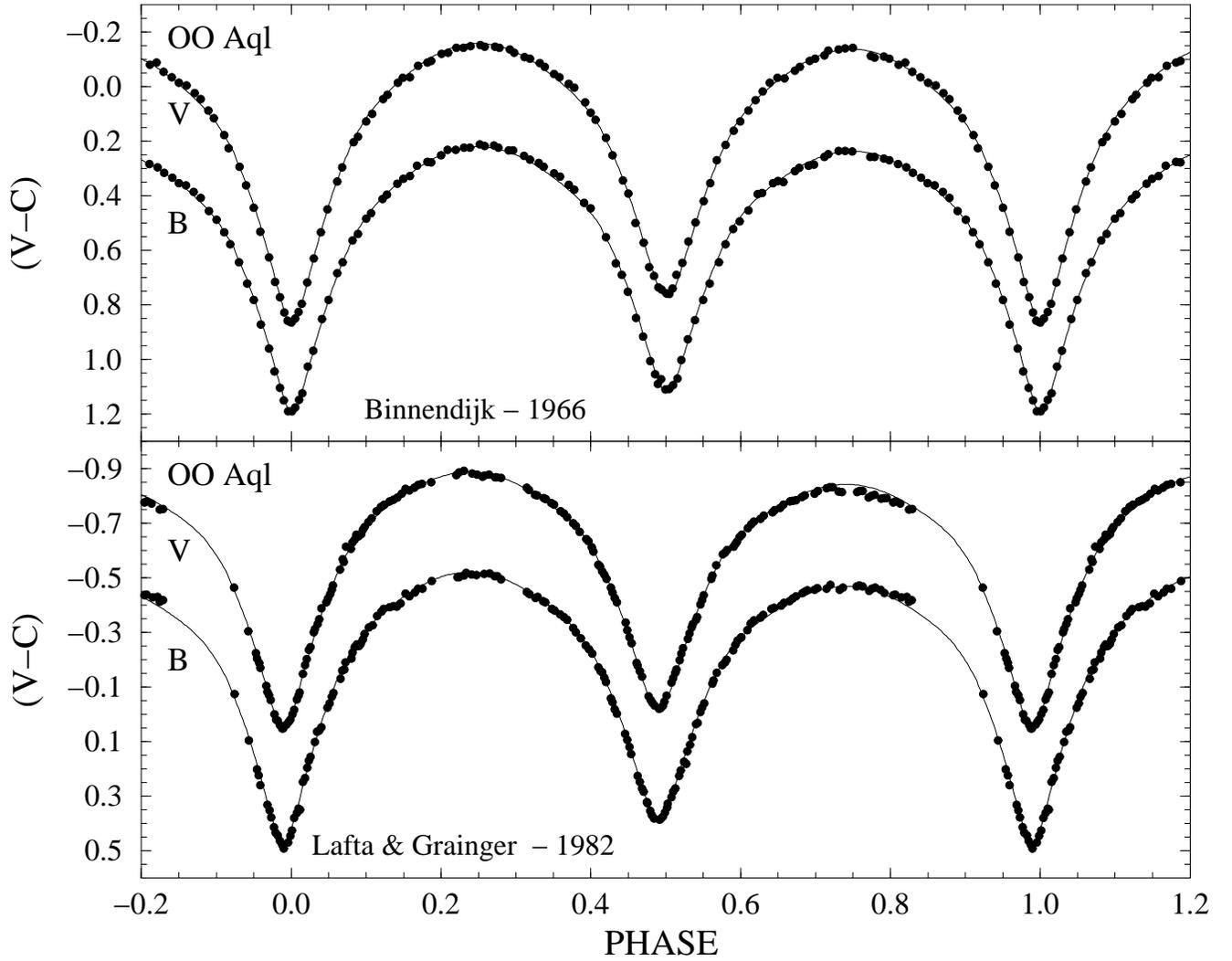}
\figcaption[]
{Observed normal-point, differential (variable $-$ comparison) B and V light
curves of OO Aql from the studies of Binnendijk (1968) for 1966 and Lafta \&
Grainger (1985) for 1982, fitted by our synthetic light curve solutions.  (Note
that different comparison stars were used in the two studies.)
\label{f4.eps}}
\end{figure*}

\begin{figure*}
\plotone{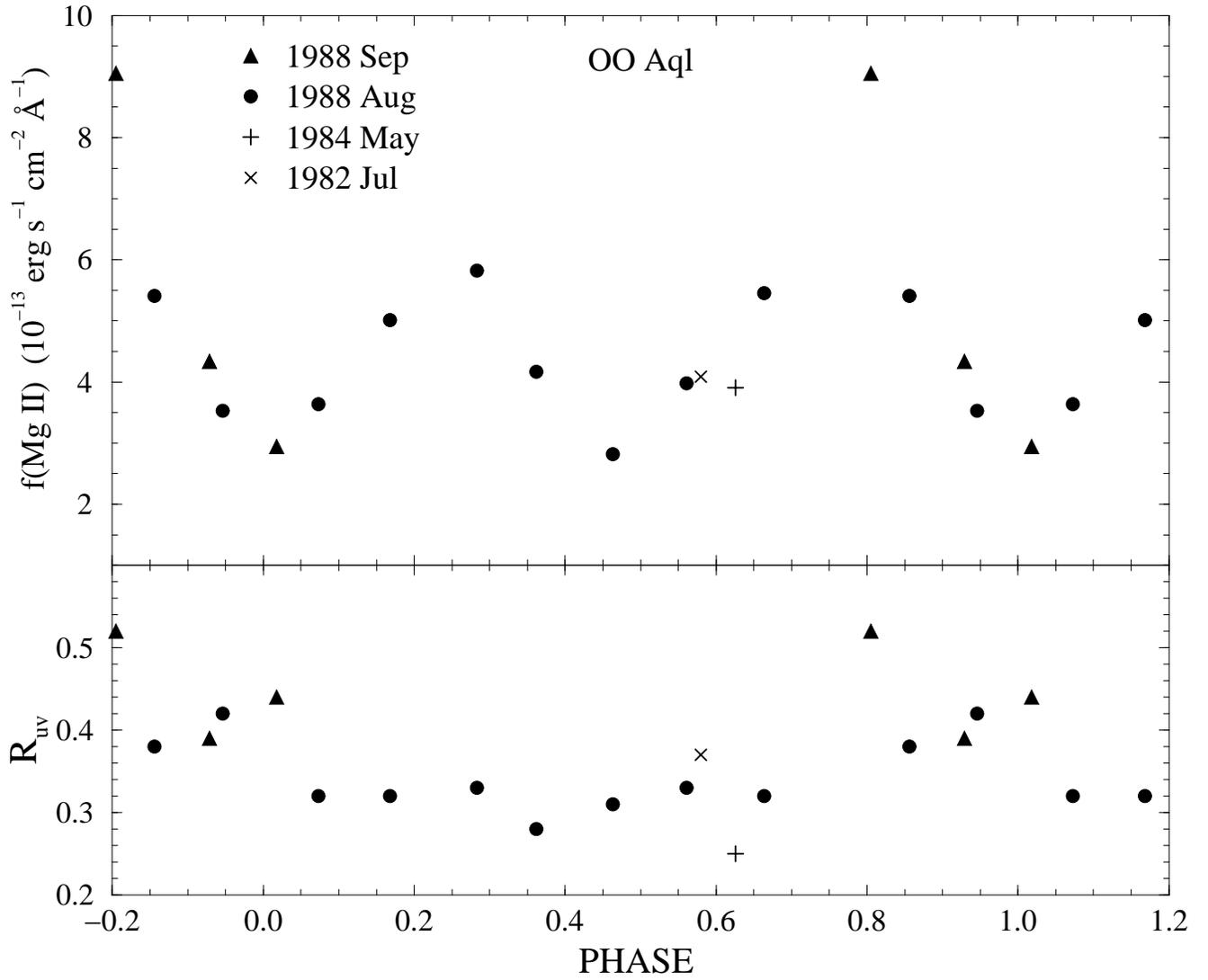}
\figcaption[]
{Plots of the Mg~{\sc ii} emission for OO Aql, (upper panel) in flux units 
(f(Mg~{\sc ii})), and (lower panel) normalized relative to the adjacent 
ultraviolet continuum (R$_{uv}$).  
Different symbols are used to distinguish the different dates: 
1988 Aug (filled circles), 1988 Sep (filled triangles), 1984 May (plus sign),
1982 Jul (cross sign).
\label{f5.eps}}
\end{figure*}

\end{document}